# An Application of Bayesian classification to Interval Encoded Temporal mining with prioritized items

C.Balasubramanian
Department of Computer Science and Engineering
K.S.Rangasamy College of Technology
Namakkal-Dt., Tamilnadu, India
rc.balasubramanian@gmail.com

Dr.K.Duraiswamy
Department of Computer Science and Engineering
K.S.Rangasamy College of Technology
Namakkal-Dt., Tamilnadu, India
drkduraiswamy@rediffmail.com

*Abstract* - **In real life, media information has time attributes either implicitly or explicitly known as temporal data. This paper investigates the usefulness of applying Bayesian classification to an interval encoded temporal database with prioritized items. The proposed method performs temporal mining by encoding the database with weighted items which prioritizes the items according to their importance from the user's perspective. Naïve Bayesian classification helps in making the resulting temporal rules more effective. The proposed priority based temporal mining (PBTM) method added with classification aids in solving problems in a well informed and systematic manner. The experimental results are obtained from the complaints database of the telecommunications system, which shows the feasibility of this method of classification based temporal mining.**

## I. INTRODUCTION

The use of maintaining large databases is in the meaningful information extracted from them. Databases are large collections of transactions present in organizations. Data mining deals with retrieving valuable information from enormous amount of data stored in such databases [1]. These transactions may have occurred at varied time points. Mining giving importance to the time at which transactions take place is called temporal mining. Temporal mining is to cluster the data based on time and then determine the association rules [2].

A temporal database consists of transactions of the form:

<Transaction-ID, items, valid time>

A temporal association rule is a binary form (AR, TimeExp), in which the left-hand side element "AR" is an association rule expressed as,

*Corresponding author .Tel. +91 4288 274741; fax: +91 4288 274745, E-mail addresses: rc.balsubramanian@gmail.com (C.Balasubramanian),

$X \Rightarrow Y$, $X \subset I$, $Y \subset I$, $X \cap Y = \varnothing$

Where "TimeExp" is a time expression [3].

aData mining especially association rule discovery tries to find interesting patterns from databases that represent the meaningful relationships between items. Association rule mining applied to large databases consumes more time because every transaction has to be scanned atleast once for any mining method which is applied. As a solution to this problem, an encoding method is considered, which will reduce the size of the database and hence the processing time required for mining becomes less [4].

In this paper, a method of encoding is proposed for a temporal database and thereafter temporal mining is performed using the Priority Based Temporal Mining method. In this method encoding is performed based on the valid time and on the weight assigned to the items in the particular transaction. This minimizes the amount of data processed while the database is scanned for association rules mining. Association rule mining giving varying importance to the different items of the transactions is called weighted mining [5]. Priority based temporal mining involves mining based on the weights assigned to the items in the transaction according to their importance and based on the time at which the transaction had taken place. This method gives better results in terms of time and computation complexity.

Naïve Bayesian classification assumes class conditional independence to simplify the computations involved [6]. In the proposed method this is used to make the resulting temporal association rules more effective in finding solutions.

The rest of the paper is organized as follows. Section 2 gives a summary of the research works carried out in areas related to temporal mining and classification. Section 3 describes the proposed classification based temporal mining methodology. Section 4 briefs the method of data preparation, which involves defining the valid time interval for encoding and temporal mining. Section 5 presents the explanation of the encoding method, which involves the merging concept. Section 6 explains how frequent itemsets are generated using the method under consideration. Section 7 presents the process of association rule mining from an encoded temporal database involving the expansion concept. Section 8 outlines the application of naïve Bayesian classification to temporal mining. Section 9 gives the performance of the above mentioned method when applied on the complaints database of a telecommunications system. Finally, in Section 10 the conclusion and future work are stated, which includes the best features of the described method and the ways in which it can be further improved.

## II. RELATED RESEARCH WORKS

Association rule identification is an integral part of any data mining process. An association is said to be present among the items if the presence of some items also means the presence of some other items [7]. Several mining algorithms have been proposed to find the association rules from the transactions [7][8][9][10]. The large itemsets were identified to find the association rules. First, the itemsets which satisfy the predefined minimum support were identified and these were called the large itemsets. Then, the association rules were identified from these itemsets. The association rules which satisfy the predefined minimum confidence were the association rules produced as the output [7]. Also, in the Graph-Based approach for discovering various types of association rules based on large itemsets, the database was scanned once to construct an association graph and then the graph was traversed to generate all large itemsets [11]. This method avoids making multiple passes over the database.

In addition to the above mentioned method of association rule mining, which overlooks time components that are usually attached to transactions in databases, the concept of temporal mining was proposed giving importance to the time constraints [3]. The concept of valid time was used to find out the time interval during which a transaction is active. Time interval expansion and mergence was performed which gives importance to the time at which a transaction had taken place, before the application of the graph mining algorithm [11] to identify the temporal association rules. For discovering association rules from temporal databases [12], the enumeration operation of the temporal relational algebra was used to prepare the data. The incremental association rule mining technique was applied to a series of datasets obtained over consecutive time intervals to observe the changes in association rules and their statistics over the time. Temporal issues of association rules was addressed with the corresponding algorithms, language and system [13][14] for discovering temporal association rules.

Further, to mine rules based on the priority assigned to the elements, weighted mining was proposed to reflect the varying importance of different items [5]. Each item was attached a weight, which was a number given by users. Weighted support and weighted confidence was then defined to determine interesting association rules.

In general, a fuzzy approach leads to results which are similar to human reasoning. A fuzzy approach involving the enhancement over AprioriTid algorithm was identified, which had the advantage of reduced computational time [15]. Also, in the mining of fuzzy weighted association rules [16], great importance had been given to the fuzzy mining concept.

Application of supervised and unsupervised learning approaches and a study of different machine learning algorithms for classification help in applying classification to association rule mining. According to the approach in [17], a user account had been split into normal and fraudulent activities using a detailed day-by-day characterization. The area under the curve was used as the statistic that exhibits the classification performance of each case. Accumulated in time, characteristics of a user yield discriminated results. The approach in [18] begins by studying the content of a large collection of emails which have already been classified as spam or legitimate email. The four machine learning methods for anti-spam filtering are discussed. An empirical evaluation for them is based on the benchmark of spam filtering. The approach in [19] predicts the itemset. In classifying a test object, the procedure uses a simple approach, which states that the first rule in the set of ranked rules that matches the test object condition classifies.This paper proposes a classification based encoded weighted temporal mining method. This identifies temporal association rules from an encoded temporal database with weighted items. The temporal database consists of transactions with their corresponding valid time intervals. The proposed method classifies the resulting association rules in such a way that the methodology for obtaining the solution is made easy.

## III. THE PROPOSED CLASSIFICATION BASED TEMPORAL MINING METHODOLOGY

The proposed methodology for classification based temporal mining is depicted in Fig.1 as follows

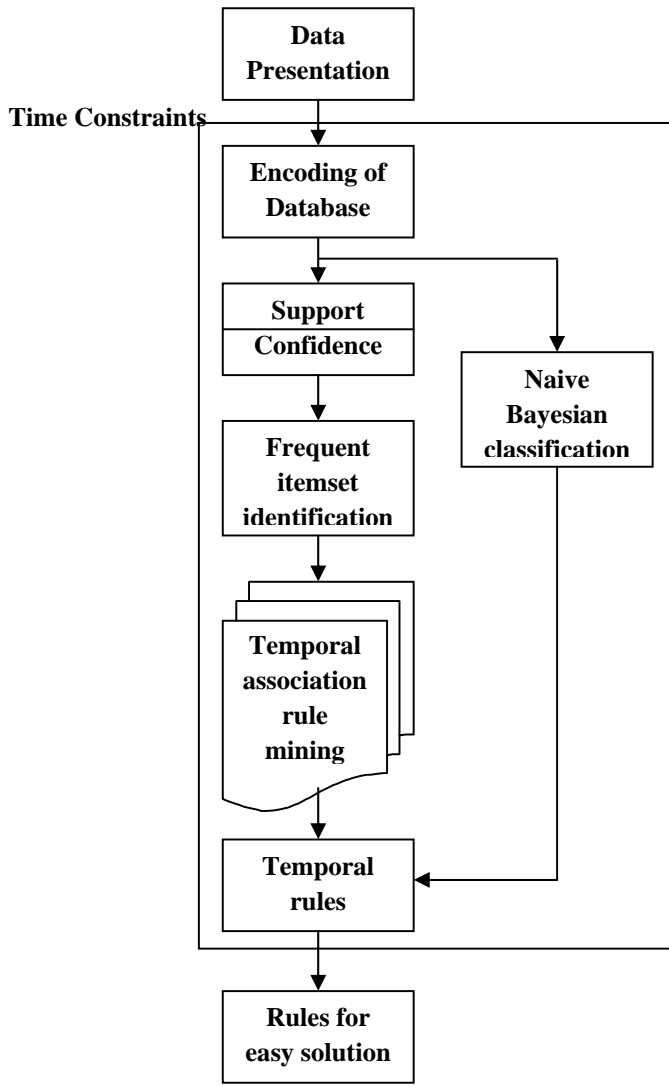

Fig.1.The proposed classification based temporal mining methodology

The proposed methodology for classification based temporal mining has the following steps.

1. **Data preparation**: It describes the way in which the time intervals are set up. The choice of the time depends upon the user and the application.
2. **Encoding of database**: This step encodes the temporal database for the given time in such a way that redundancy is avoided by merging.
3. **Frequent itemset generation**: This identifies the large itemsets by applying the priority based temporal mining method.
4. **Temporal association rule mining**: It identifies the association rules in each interval and applies transitive property to find out the relationship between the associations rules resulting from different time intervals. It also expands the time intervals if the time intervals are continuous and hold the same association rules which satisfy the minimum confidence.
5. **Naïve Bayesian classification**: It identifies the probability for the classes involved which leads to temporal rules for easy solutions.

## IV. DATA PREPARATION

In this step, depending upon the user's choice and the application under consideration, the length of the interval during which data is accumulated, and the number of such intervals is decided. Then, the subset of the transactions in the temporal database which conform to each of the time interval is extracted. The application under consideration is the addressing of complaints from a complaints database of the telecommunications department.

## V. THE ENCODING OF DATABASE

The most common approach to interval encoding of temporal databases is to use intervals as codes for one dimensional set of time instants[3][12]. The choice of this representation is based on the following empirical observation: Sets of time instants describing the validity of a particular fact in the real world can be often described by an interval or a finite union of intervals. For simplicity, a discrete integer-like structure of time is assumed. However, dense time can also be accommodated by introducing open intervals.

Let Interval-based Domain be TI and let TP = (T,<) be a discrete linearly ordered point-based temporal domain. The set I (T) is defined as

$I(T) = \{(a, b) : a \leq b, a \in T \cup \{-\infty\}, b \in T \cup \{\infty\}\}$

where < is the order over TP extended with $\{(-\infty, a), (a, \infty),$

$(-\infty, \infty): a \in T\}$. The elements of I(T) are denoted by [a,b] which is the usual notation for intervals. The four relations on the elements of I(T) denoted by [a,b] are defined as follows:

$([a, b] < -- [a', b']) \Leftrightarrow a < a'$

([a, b] < +- [a', b']) ⇔ b< a'

([a, b] < -+ [a', b']) ⇔ a< b'

([a, b] < ++ [a', b']) ⇔ b< b'

for [a, b], [a', b'] ∈ I(T).

The structure TI = (I(T), <--, < +- , <-+, <++) is the interval based temporal domain corresponding to TP.

A concrete (timestamp) temporal database is defined analogously to the abstract (timestamp) temporal database. The only difference is that the temporal attributes range over intervals (TI) rather than over the individual time instants (TP).

For the priority based temporal mining method the two levels of encoding are encoding based on the valid time, and encoding based on the weight assigned to the items in the transaction under consideration. Every complaint is assigned a unique number (weight) which denotes the priority of the complaint in terms of its critical nature. For example: A-0.1, B-0.2, C-0.3, D-0.4, E-0.5 and so on. The encoding of the database for this method is as follows in table 1 and table 2.

TABLE 1: ENCODED DATABASE AT VALID TIME INTERVAL D1

| Tid | Complaint weights W | Count | Weighted Support =∑Wx(Count+1) |
|---|---|---|---|
| 01 | 0.2 | 3 | 0.6 |
| 02 | 0.1 | 2 | 0.2 |
| 03 | 0.3 | 1 | 0.3 |

TABLE 2: ENCODED DATABASE AT VALID TIME INTERVAL D2

| Tid | Complaint weights W | Count | Weighted Support =∑Wx(Count+1) |
|---|---|---|---|
| 06 | 0.1 | 2 | 0.2 |
| 07 | 0.5 | 4 | 2.0 |
| 09 | 0.4 | 3 | 1.2 |

The count field is zero if the set of complaint codes occurs only once. If the same set of complaints occurs more than once then merging takes place i.e. the count field is incremented by 1 for each repetition and the set of complaints appears only once in the database. Thus redundancy is avoided. The value in the count field +1 gives the support value for the specific set of complaints which is compared with the predefined minimum support value to identify the large itemsets. In this method, the weight of the complaint is given instead of the code according to the priority. The candidate itemsets satisfying bounded support are checked for their weighted support. The weighted support is calculated as the product of the summation of the weights and the corresponding count+1 value for each large itemset (obtained from the table). This weighted support is compared with the weighted minimum support value to identify the large itemsets.

## VI. FREQUENT ITEMSET GENERATION

An association rule that strictly satisfies both minimum support threshold and minimum confidence threshold is called as a strong association rule. Similarly, a strong temporal association rule is defined as follows. Let min_s and min_c represent minimum support threshold and minimum confidence threshold respectively. If and only if support ≥ min_s, and confidence ≥ min_c, during [$t_s$, $t_e$], rule X⇒Y is a temporal association rule, which could be described as,

X⇒Y (support, confidence, [$t_s$, $t_e$])

Itemset means the collection of items. If there are k items, it is called as k-itemset. The itemset that satisfies min_s is called frequent itemset.

Given a database with T transactions belonging to a specified duration [$t_s$, $t_e$], the bounded support (BS) of a large k-itemset X is defined to be the transaction number containing X within the specified valid time, and it must satisfy eqn(1) given by:

$$BS(X) = \frac{(\sum_{i=1}^{T} Count_i + T) \times wmnspt}{\sum_{\forall i \in X} W_i} \quad (1)$$

where $W_i$ is the summation of the weights of all the items in large k-itemset X in a specific duration [$t_s$, $t_e$]. The value of count is taken from the table and its summation for a particular duration of time added with T gives the total number of transactions in the specific time duration. The threshold value for the weighted minimum support is denoted by wmnspt which is defined by the user. The bounded support value according to eqn(1) is calculated for each large k-itemset so that the itemset which are not necessary for further calculations can be avoided. The bounded support calculations are done so that the necessary pruning may be performed and time may be saved. The weighted minimum support is defined by eqn (2) as,

$$WS(X) = \sum_{\forall i \in X} W_i \times (Count + 1) \qquad (2)$$

where $\sum_{\forall i \in X} W_i$ is the sum of the weights of the items present in itemset X in a given valid time $[t_s, t_e]$. Count + 1 gives the number of transactions containing the specific itemset within the above mentioned time interval $[t_s, t_e]$.

The large k-itemsets obtained as a result of pruning will be checked for its weighted support as follows: If the weighted support of an itemset is greater than the threshold value of the weighted minimum support, i.e. if weighted support ≥ wmnspt, then it is considered as a candidate for the next step.

Thus the large itemsets for a specific time interval, for transactions with prioritized items, have been identified and these are the itemsets which will be used for generating the association rules by using the minimum confidence threshold value denoted as min_c.

## VII. TEMPORAL ASSOCIATION RULE MINING

Association rules are generated from the large itemsets which satisfy the user defined minimum confidence min_c. The confidence of the association rule X⇒Y is the probability that Y exists given that a transaction contains X and is given by eqn(3) as,

$$Pr(Y \mid X) = \frac{Pr(X \cup Y)}{Pr(X)} \qquad (3)$$

In large databases, the support of X⇒Y is taken as the fraction of transactions that contain X∪Y. The confidence of X⇒Y is the number of transactions containing both X and Y divided by the number of transactions containing X.

In the case of priority based temporal mining, the confidence value is represented by eqn(4) as,

$$Confidence = \frac{Weighted\ Support(X \cup Y)}{Weighted\ Support(X)} \qquad (4)$$

The weighted support value is obtained from the table. This confidence value has to be greater than or equal to the predefined minimum confidence threshold for the corresponding large itemset to be included as an association rule in the output.

The property of transitivity and the concept of expansion of time intervals are considered to show the relationship between the different valid time intervals taking into account the resulting association rules from each of the time intervals. This new concept gives importance to the valid time interval as well as to the relationship between the time intervals. The temporal association rules produced are found to be highly related with the time constraints involved.

## VIII. NAÏVE BAYESIAN CLASSIFICATION

Classification is one of the most typical operations in supervised learning, but hasn't deserved much attention in temporal data mining. In fact, a comprehensive search of applications based on classification has returned relatively few instances of actual uses of temporal information. Since traditional classification algorithms are difficult to apply to sequential examples, an interesting improvement consists on applying a pre-processing mechanism to extract relevant features. One approach is idea consists on discovering frequent subsequences, and then using them, as the relevant features to classify sequences with traditional methods, like Winnow.

Classification is relatively straightforward if generative models are employed to model the temporal data. Deterministic and probabilistic models can be applied in a straightforward way to perform classification. The Naïve Bayesian classification is based on probabilities represented by P and assumes class conditional independence. That is, it presumes that the values of the attributes are conditionally independent of one another given the class label of the tuple. The naïve Bayesian classifier predicts that tuple X belongs to the class $C_i$ if and only if

$P(C_i \mid X) > P(C_j \mid X)$ for $1 \leq j \leq m, j \neq i$

The class $C_i$ for which $P(C_i \mid X)$ is maximized is called the maximum posteriori hypothesis (eqn (5))

$$P(C_i \mid X) = \frac{P(X \mid C_i) P(C_i)}{P(X)} \qquad (5)$$

No dependence relationships among the attributes is emphasized by eqn (6) given below.

$$P(X\backslash C_i) = \prod_{i=1}^{n} P(x_k \backslash C_i) \qquad (6)$$

Using the above mentioned classification, the resulting temporal rules are made effective and it paves the way for easy solving methodologies.

## IX. PERFORMANCE EVALUATION

The experimental results for the proposed classification based temporal mining methodology are obtained by considering a temporal database of 25,000 records which are transactions containing the complaints from customers with different valid time intervals. The complaints in each transaction are assigned priority by assigning unique number to each complaint according to the importance of the complaint from the user's perspective. This large database is encoded by using interval encoding of temporal databases.

Database encoding has been applied to a static database prior to the application of Apriori or anti Apriori. To make it scalable, the same has been applied to a dynamic database, which involves time constraints. Considering the telecommunication temporal database, which addresses complaints, performance of the Apriori family of algorithms and the Anti-Apriori algorithm is as given in Fig 2.

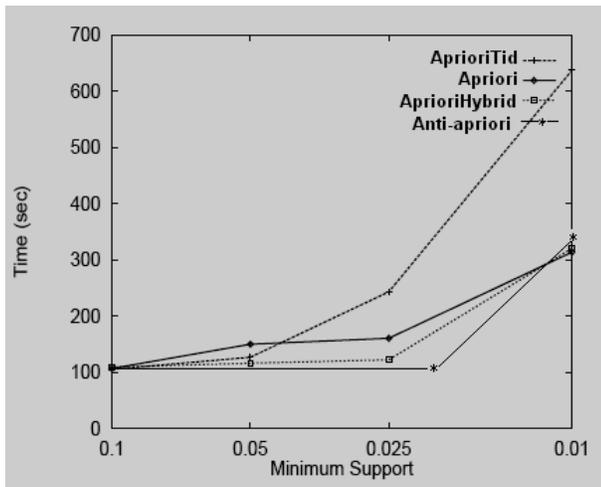

Fig.2.Performance of Apriori family algorithms

Fig3 below shows the usage of memory before and after encoding. It is also found that the encoding method leads to faster generation of temporal association rules.

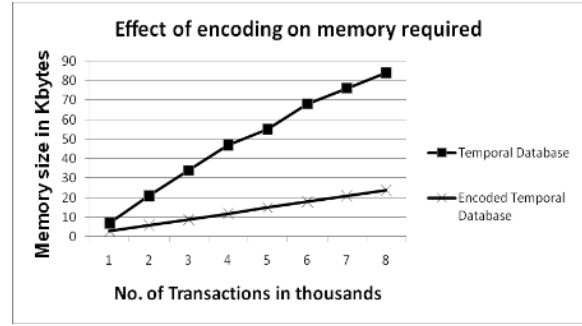

Fig. 3.Memory usage graph

The classification based temporal mining methodology has better performance in terms of the logic used. The performance of this method increases with increase in the number of tuples within a given time interval. The computational complexity decreases with increase in the number of transactions. This is depicted in fig4 as follows.

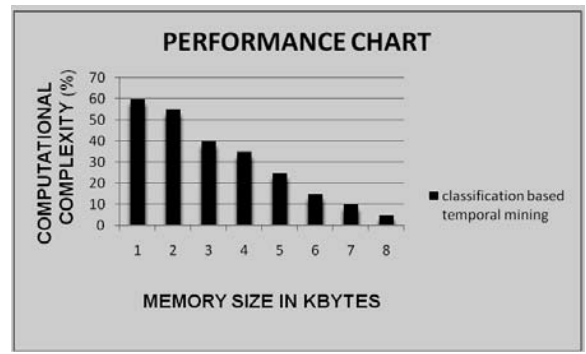

Fig.4.Performance of classification based temporal mining

## X. CONCLUSION AND FUTURE ENHANCEMENTS

Classification based temporal mining which involves assigning priorities, always leads to more advantages than other concepts which treat all items uniformly since prioritizing reduces the time involved. In a similar manner prioritizing the items and then encoding the temporal database has lead to lesser complexities of time and computation. Introduction of Bayesian classification in temporal mining has lead to more effective temporal rules. A very important fact is that these results are obtained by including the time constraints (i.e.) within the specified valid time interval. All applications that are time based need to satisfy the real time constraints. Hence applying classification based temporal mining which involves the important concepts of priority, encoding, valid time interval, temporal mining and classification to real time applications which has response time constraints will improve the performance measures in a sharp and distinct manner.

This work may be further extended to improve the performance of systems that involve real time data in the form of audio, video and other multimedia objects which are stored as data items in a database with valid time constraints i.e. a temporal database of transactions containing complex form of data. Also classification tools can be used to categorize the results obtained in order to identify the methodology of solving problems in a systematic way with less complexity.